# EMITTANCE MEASUREMENTS WITH WIRE SCANNERS IN THE FERMILAB SIDE-COUPLED LINAC *

E. Chen, R. Sharankova, A. Shemyakin, J. Stanton
Fermi National Accelerator Laboratory, Batavia, USA

## Abstract

The Fermilab Side-Coupled Linac accelerates H$^-$ beam from 116 MeV to 400 MeV through seven 805 MHz modules. Twelve wire scanners are present in the Side Coupled Linac and four are present in the transfer line between the Linac and the Booster synchrotron ring. These wire scanners act as important diagnostic instruments to directly collect information on the beam's transverse distribution. The manipulation of the conditions of wire scanner data collection enables further characterization of the beamline, such as calculating emittance and the Twiss parameters of the beam at select regions, which we present here.

## INTRODUCTION

The Fermilab Side-Coupled Linac accelerates 116.5 MeV H$^-$ beam to 400 MeV via 805 MHz side-coupled cavities [1]. The cavities are organized into seven main modules (Modules 1-7) with four sections in each module, as well as a transition section (Module 0) composed of 805 MHz buncher (B) and Vernier (V) cavities, Fig. 1.

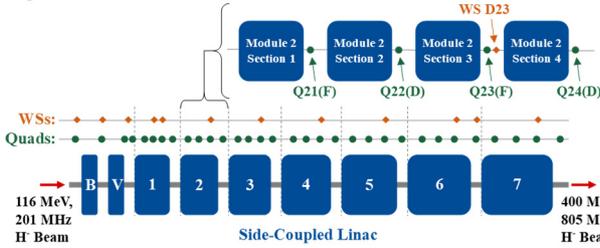

Figure 1: Schematic of the Fermilab Side-Coupled Linac and representative layout inside Module 2. Naming convention for diagnostic instruments follows the structure of Qmn (quads, green dots) or Dmn (WSs, orange diamonds), where m = module number and n = section number.

Diagnostic, steering, and focusing elements are present in between sections and modules, including an alternating FODO lattice of quadrupoles [1] and 12 wire scanners (WSs) throughout the Side-Coupled Linac [2]. The WSs yield transverse profiles in the X and Y directions (Fig. 2), as well as in a 45° direction not discussed here.

The abundance of these diagnostic devices in the Side-Coupled Linac facilitates the use of WSs to measure the transverse emittance and the Twiss parameters of the H$^-$ beam. Such measurements have not been performed in more than a decade [2]; however, we seek to experimentally characterize the present state of the Fermilab Linac. Here we present the results of the recent study analyzing transverse beam properties in the Side-Coupled Linac.

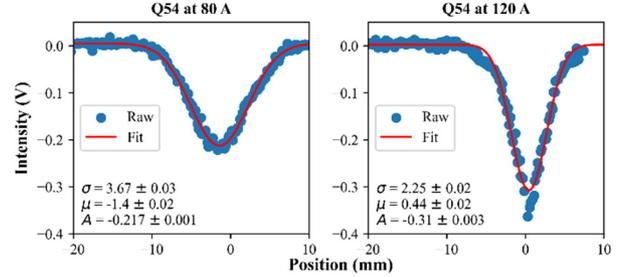

Figure 2: Example of vertical (Y) wire scans from quadrupole scan Q54-D63 with Q54 at 80 A and 120 A.

## MEASUREMENTS OF EMITTANCE AND TWISS PARAMETERS

The transverse emittance and Twiss parameters were measured via two methods: a quadrupole scan and a beam width measurement at several locations in a drift space.

To reduce the duration of the wire scans, the scan procedure was modified to run at a constant speed. Previously, the wire frame was moved in steps, requiring significant time to equilibrate at each position and an overall scan time of 10-15 minutes. This long scan time is incompatible with a quadrupole scan, in which a large number of wire scans is required. Instead, the modified procedure collects position and intensity data on each beam pulse while moving the wires at a constant speed across the beam pipe, reducing scan time tenfold to 1.25 minutes and demonstrating comparable data quality [3]. The step in space between the points is defined by the wire frame speed and pulse frequency. To address the low signal-to-noise ratio, the beam profile width is characterized by the standard deviation of the fitted Gaussian function $\sigma_{RMS}$. The use of beam width from the Gaussian fit additionally allows inclusion of scans with incomplete beam profiles due to insufficient scan range. The RMS errors associated with the Gaussian fit are assigned to all points shown in Figs. 3 and 4. A Python program was written to schedule the wire scans and vary the quadrupole current for the quadrupole scan, enabling faster collection.

Data for both methods were collected in July 2024 with 18.7 mA beam at the output of the Linac and 5 Hz pulse frequency. For the quadrupole scan, wire scans taken when moving the frame both "in" and "out" of the beam pipe with a 5 μs pulse length. For the propagation in free space experiment, data were taken in both the "in" and "out" directions and averaged.

### Quadrupole Scan

In the thin-lens approximation, the quadrupole scan is defined by the characteristic equation in Eq. (1).

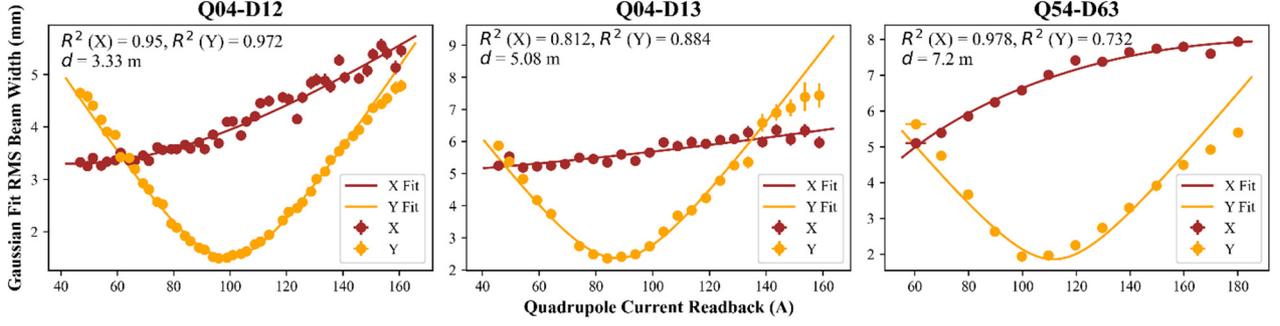

Figure 3: Quadrupole scan results in the X and Y planes from quadrupole-WS pairs Q04-D12 (A), Q04-D13 (B), and Q54-D63 (C). In the figures, the parabolic fit result was scaled to be displayed relative to the raw RMS beam width.

$$\sigma_{11,f} = (d^2\sigma_{11,i})(kl)^2 - (2d\sigma_{11,i} + 2d^2\sigma_{12,i})(kl) \\ + (\sigma_{11,i} + 2d\sigma_{12,i} + d^2\sigma_{22,i}) \quad (1)$$

In Eq. (1), $\sigma_{nm}$ refers to elements of the beam distribution matrix or σ-matrix [4] at the quadrupole (initial, $i$) and WS (final, $f$); $d$ refers to the distance of the drift between the quadrupole and WS; and $kl$ refers to the inverse focal length of the quadrupole proportional to its current.

The emittance $\varepsilon$ (RMS, geometric) is related to the $\sigma_{11}$ term of the σ-matrix via Twiss β-function as $\sigma_{11} = \varepsilon\beta = \sigma_{RMS}^2$. Thus, the emittance and Twiss parameters can be found from a parabolic fit of the squared RMS beam width against the inverse focal length.

Three quadrupole scans were executed, assessing beam parameters at two different positions. Intermediate accelerating and focusing elements were turned off for the scan. The first position is at the end of the transition section, assessed with quadrupole Q04 and two different downstream WSs, D12 and D13. The second position is at the end of Module 5, assessed with quadrupole Q54 and WS D63.

The quadrupole scan data and fit are shown in Fig. 3. The calculated emittance and Twiss parameters α, β, and γ are summarized in Table 1. Note that the shape of the curves for X data in scans Q04-D13 and Q54-D63 does not allow reconstruction with Eq. (1).

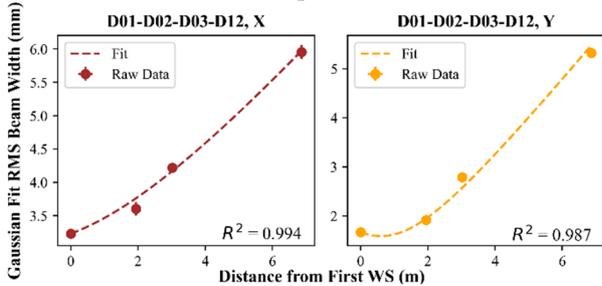

Figure 4: Results of propagation in free space in X and Y planes. In the figures, the fit result was scaled to be displayed relative to the raw collected data of RMS beam widths of the downstream WSs.

*Propagation in Free Space*

The propagation in free space study is defined by the characteristic equation in Eq. (2).

$$(\sigma_{11,f} - \sigma_{11,i}) = (\sigma_{22,i})(d_f)^2 - (2\sigma_{12,i})(d_f) \quad (2)$$

In Eq. (2), $\sigma_{nm}$ refers to elements of the σ-matrix at the first upstream WS (initial, $i$) and all downstream WSs' (final, $f$) positions; and $d_f$ refers to the distance of the drift between the first upstream WS and each downstream WS.

Using Eq. (2), the emittance and Twiss parameters can be found from a two-parameter parabolic fit of the beam widths measured by the downstream WSs vs. the drift distances.

The propagation in free space measurement was performed in the transition section where the highest concentration of WSs occurs in the Linac, using four WSs (D01-D02-D03-D12). The data and fits are shown in Fig. 4. Transverse characteristics of the beam from this experiment are presented in Table 1.

For both measurements, the errors indicated in Table 1 are calculated from the statistics of the parabolic fit. The full covariance matrix from the parabolic fit is propagated with the Jacobian [5] to yield the final calculations for emittance and Twiss parameters. The parabolic fit accounts for errors from individual wire scans via use of the SciPy curve_fit() function's sigma and absolute_sigma options [6].

## CONSISTENCY OF EMITTANCE DATA

The results summarized in Table 1 assess three locations in the Side-Coupled Linac, where the vertical plane results from Q04-D12 and Q04-D13 both probe the emittance and Twiss parameters at Q04. Directly comparing these two quadrupole scans shows the results are consistent within one standard deviation of each other, except for β which is within four standard deviations.

The normalized emittance is expected to remain conserved throughout the Linac. The two values for the horizontal emittances are consistent, primarily because of the large error associated with the free-space propagation result. However, the derived values of the vertical emittance at three different locations differ by several standard deviations. Several sources may contribute to this discrepancy.

Uncertainty exists in the values used for the distances between beamline elements and the quadrupole calibration constants. Recent measurements of distances led to the use of a positioning error of ± 4 cm RMS in the error calculation for Table 1, while quadrupole calibrations were taken from the original design, and their errors were assumed negligible.

Table 1: Emittance and Twiss Parameters in the Side-Coupled Linac

| Scan | Direction | ε RMS, normalized (mm-mrad) | α | β (m) | γ (m⁻¹) | Energy (MeV) |
|---|---|---|---|---|---|---|
| D01-D02-D03-D12 | Horizontal | 0.9 ± 0.1 | -0.3 ± 0.1 | 5.7 ± 0.7 | 0.20 ± 0.01 | 116.5 |
| D01-D02-D03-D12 | Vertical | 0.68 ± 0.01 | 0.32 ± 0.03 | 2.09 ± 0.04 | 0.53 ± 0.02 | 116.5 |
| Q04-D12 | Horizontal | 0.83 ± 0.04 | -0.01 ± 0.05 | 1.64 ± 0.05 | 0.61 ± 0.02 | 116.5 |
| Q04-D12 | Vertical | 0.80 ± 0.02 | -3.07 ± 0.04 | 7.59 ± 0.06 | 1.37 ± 0.02 | 116.5 |
| Q04-D13 | Vertical | 0.78 ± 0.01 | -2.98 ± 0.05 | 7.0 ± 0.1 | 1.41 ± 0.02 | 116.5 |
| Q54-D63 | Vertical | 0.700 ± 0.009 | -3.81 ± 0.04 | 11.9 ± 0.1 | 1.30 ± 0.01 | 313.6 |

Another possible contributor to the discrepancy is the use of the Gaussian fit to characterize the beam width. The procedures of Eq. (1) and (2) are strictly applicable to the second momenta of the beam distribution. In these measurements, they are replaced by Gaussian fits as the low signal-to-noise of the wire scanner data precludes direct calculation of RMS beam width. Although our data are generally well characterized by a Gaussian fit, Fig. 2, there are long-standing concerns about the non-Gaussian nature of the beam in the Linac [7]. In the non-Gaussian examples, there is additional peakedness and minor beam shoulders. By visual inspection, this predominantly occurs in the high current data of the Q54-D63 scan.

However, the main source of the discrepancy in Table 1 is likely related to the inclusion of wire scans where the beam is too large, which occur at extreme current values. With large beam widths, part of the beam may be scraped by the 30 mm aperture of Linac [1]. Although there was no effect to the smoothness of the beam profiles at ±15 mm, the beam scraping is confirmed by elevated loss monitor signals. Additionally, high loss data corresponded to a reduction of transverse profile area from the Gaussian fits, up to 30% in the high current data of Q54-D63.

The unphysical shape of the X curve for the Q54-D63 series in Fig. 3 may originate from this loss. Losses in the X plane may additionally affect the orthogonal plane via secondary particles or if the beam distribution is coupled.

If these high-loss data are included, the beam width is reconstructed from a profile with larger errors and introduces systematic errors into the parabolic fit that are not characterized by the statistical errors reported in Table 1. If excluded, the range of scans for the parabolic fit may become insufficient.

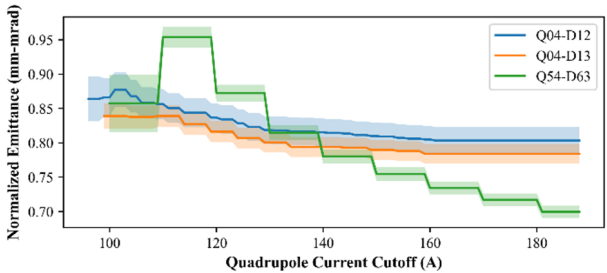

Figure 5: Normalized emittance of quadrupole scans when data above a quadrupole current value is excluded. Shaded regions correspond to the error of the emittance.

A test of this hypothesis is presented in Fig. 5, where the fitting procedure was applied to subsets of the data restricted to the quadrupole current below a cut. The results show that data removal significantly affects the emittance in the case of Q54-D63 where emittances with different cuts span from 0.700 to 0.95 mm-mrad. In contrast, the cut data for Q04-D12 and Q04-D13 produce self-consistent (within one standard deviation) emittance values for cuts above 115A and 120 A, respectively. These results make sense in context of the worse fit of Q54-D64, Fig. 3.

The results of the exclusion test emphasize the systematic effect of beam loss on the quality of the quadrupole scan, which is not well-characterized by the statistical errors from the parabolic fit or the initial Gaussian fit of the data. This is similarly relevant for the free space propagation experiment, suggesting that the beam width at the farthest wire scanner D12 becomes less reliable. The normalized RMS emittance in the Side-Coupled Linac can thus be described as 0.8 ± 0.1 mm in both horizontal and vertical planes.

The emittance measured in the Side-Coupled Linac has not been experimentally measured and reported in over a decade [2]. However, the emittance can be compared with measurements in other locations. At the Linac injection, RMS emittance measurements of the 750 keV beam with emittance probes varied widely within a year, between 0.5-0.9 mm-mrad and 0.5-1.5 mm-mrad for the horizontal and vertical planes, respectively [8]. Unfortunately, no emittance measurements there were made concurrently with those presented in this report. The Fermilab Booster downstream (2021, via ionization profile monitors) observes RMS emittance of the injected Linac H- beam of 1 ± 0.2 mm-mrad [9]. Therefore, no clear discrepancy is found.

## CONCLUSION

The transverse properties of the H- beam in the Fermilab Side-Coupled Linac were measured with quadrupole scans and free-space propagation. The resulting RMS emittance is 0.8 ± 0.1 mm in both horizontal and vertical planes. From here, we seek to refine the RMS beam width calculation to better characterize profiles with non-Gaussian features, in addition to developing a method to characterize points unsuitable for inclusion in a quadrupole scan or free space propagation experiment.